\documentclass{phbauth}        % Don't modify
\usepackage{graphicx}
\usepackage{amssymb}

\begin{document}

\begin{frontmatter}

% Use lower case letters in the title.
\title{Disorder effects on transport 
near AFM quantum phase transitions}

\author[address1]{Achim Rosch\thanksref{thank1}}
%\author[address1]{xxx},
%\author[address2]{xxx}

\address[address1]{Center for Materials Theory, 
Physics \& Astronomy, Rutgers University, Piscataway, NJ 08854, USA}
%\address[address2]{xxx}

% The corresponding author should be distinguished and his email
% address and/or fax number must be given. His mailing address has to
% be complete: the proofs are send to this address around
% January 1, 2000. The address for sending proofs has to be indicated
% as "present address", if it is different from the address above.
\thanks[thank1]{E-mail: rosch@physics.rutgers.edu}

\begin{abstract}
We discuss three different scenarios recently proposed
to account for the non-Fermi liquid behavior near antiferromagnetic (AFM)
quantum critical points in heavy-Fermion systems: 
(i) scattering of Fermi liquid
quasiparticles by strong spin fluctuations near the spin-density-wave
instability,
(ii) the breakdown of the Kondo effect due
to the competition with the RKKY interaction, and (iii) the formation of
magnetic regions due to rare configurations of the disorder.
Here we focus on the first scenario and show
that it explains in some detail  the anomalous temperature dependence of the
resistivity observed e.g.  in CePd$_2$Si$_2$, CeNi$_2$Ge$_2$ or
CeIn$_3$. The interplay of strongly anisotropic
scattering due to critical spin-fluctuations and weak isotropic
impurity scattering leads to a regime with a resistivity
 $\rho \approx \rho_0 +\Delta\rho (T)$, $\Delta\rho (T) \equiv
  T^{3/2} f(T/\rho_0)
\propto  T \sqrt{\rho_0}$ for sufficiently large $T$ and small $\rho_0$.
\end{abstract}


\end{frontmatter}

\newcommand{\w}{\omega}
\newcommand{\text}[1]{\mathrm{#1}}
\newcommand{\kf}{k_F}
\newcommand{\Q}{{\vec{Q}}}
\newcommand{\kk}{{\vec{k}}}
\newcommand{\kappaV}{{{\mbox{\boldmath $\kappa$}}}}
\newcommand{\q}{{ \vec{q}}}
\newcommand{\A}{\text{\AA}}
\newcommand{\IM}{\text{Im}}
\renewcommand{\vec}[1]{{\bf #1}}

% The main text begins here. The \section commands are optional.
\section{Introduction}

In  recent years an increasing number of heavy-fermion metals
near an AFM quantum critical point (QCP) were shown to
display striking deviations from conventional Fermi-liquid behavior
\cite{loehneysen,julian,gegenwart}. In particular,
the resistivity rises
as a function of temperature with exponents smaller than $2$,
the specific heat diverges logarithmically or shows a $\sqrt{T}$ cusp,
and the  susceptibility 
shows anomalous corrections of the form $T^{\alpha}$, with $\alpha < 1$.

The common feature of the systems we are considering is that,
as a function of some control parameter, which can be doping, pressure or
magnetic field, antiferromagnetic order is suppressed. The
anomalous behavior mentioned above occurs in the vicinity
of the 
QCP defined by the vanishing of the N\'eel temperature\cite{super}. 
We stress that both magnetic and non-magnetic phases are heavy-Fermion metals
which display, for example, the characteristically large specific heat.
It is commonly believed that the QCP
results from the competition between the Kondo screening of magnetic moments
and the AFM correlations induced by the RKKY interaction \cite{doniach}.
At least three different theoretical scenarios have been proposed to
account for the observed behavior near the QCP.

\begin{figure}[t]
%h=here, t=top, b=bottom, p=separate figure page
\begin{center}\leavevmode
\includegraphics[width=0.8\linewidth]{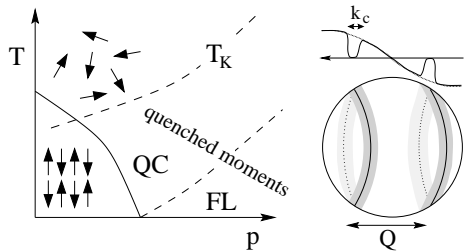}
\caption{ Scenario 1: Spin-density-wave transition in a Fermi liquid.
In a Kondo lattice, below a scale $T_K$,  heavy quasi particles describe
well the low energy excitations of the Fermi liquid. They
undergo a AFM phase transition which is
in the universality class of a SDW transition \cite{hertz}.
The QPT is driven e.g. by pressure $p$, doping or
magnetic fields.
Right figure: Near the QCP, scattering on the Fermi surface is
enhanced along ``hot lines''
with $\epsilon_{\kk}\approx \epsilon_{\kk \pm \Q}\approx \mu$, where $\Q$
is the ordering vector of the AFM. In the remaining ``cold regions''
inelastic scattering is weak.
}
\label{scen1}
\end{center}
\end{figure}

The first scenario is based on the
the assumption that in a heavy Fermion system below
a scale $T_K$  (see Fig.~\ref{scen1}) the low energy excitations are
(heavy) quasi particles and their collective
excitations. 
In this case the QCP should be in the same universality class as
the weak-coupling spin-density wave (SDW) transition in a Fermi liquid
studied by Hertz \cite{hertz}. 
More precisely, the behavior near the QCP is determined by the
mutual interaction
between Landau damped spin fluctuations and inelastically scattered 
quasi-particles.
The bulk of this paper concerns the transport properties near the
QCP, an issue of great experimental relevance.
In spite of the substantial amount of work in this area \cite{hlubina}
the interplay of strongly anisotropic inelastic scattering and
weak isotropic impurity scattering had not been properly analyzed
until recently \cite{roschRes}.

\begin{figure}[t]
\begin{center}\leavevmode
\includegraphics[width=0.52\linewidth]{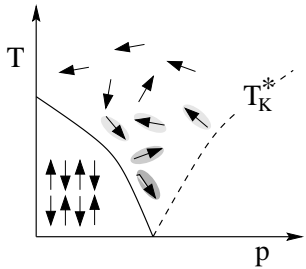}
\caption{ Scenario 2: Breakdown of the Fermi liquid. 
The Kondo effect, which tries to screen local
moments, breaks down near the QCP due to the competition with
 the RKKY interaction and singular magnetic fluctuations \cite{si,coleman}. 
An effective
Kondo temperature $T_K^*$ vanishes at the QCP.
}\label{scen2}\end{center}\end{figure}

While this first scenario appears internally consistent, it is far
from clear that the notion of heavy quasi particles survives near
the QCP. For example, one can envision a situation where the Kondo effect breaks
down near the QCP as a result of the competition of the Kondo effect
with the RKKY interaction and
singular magnetic fluctuations. In this second scenario the effective
``Kondo'' temperature $T_K^*$ vanishes at the QCP as shown in
Fig.~\ref{scen2} and the quantum critical behavior is
dominated by {\em local} magnetic fluctuations.
A microscopic justification of this scenario is still missing
but might be realized along the lines 
proposed recently by Q.~Si {\it et al} \cite{si}.
At present,
CeCu$_{6-x}$Au$_x$ seems to be the best experimental candidate 
for such a scenario \cite{schroeder,coleman}.
At the QCP, the dynamic susceptibility has been fitted to
$\chi(\vec{q},\w)^{-1}\approx f(\vec{q})+a (b T-i \w)^{\alpha}$
\cite{schroeder}. Astonishingly, the anomalous exponent $\alpha
\approx 0.8$ shows up not only near the ordering wave vector $\Q$ of
the AFM, but over the entire Brillouin zone. This suggests
a {\em local}
origin for the non-Fermi liquid behavior as might be induced by a 
breakdown of the Kondo effect, consistent with the ideas mentioned above.

\begin{figure}[t]
\begin{center}\leavevmode
\includegraphics[width=0.52\linewidth]{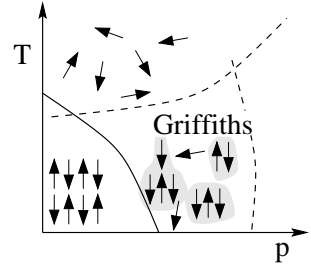}
\caption{ Scenario 3: Disorder. Due to disorder, the effective
N\`eel temperature
varies within a sample. Near the QCP these fluctuations get more
and more important and change the critical behavior. In a region around
the QCP, the thermodynamics is dominated by {\em rare} disorder configurations
which lead to the formation of magnetic domains even in the paramagnetic phase.
}\label{scen3}\end{center}\end{figure}

The third scenario I would like to mention is driven by the
presence of sufficiently strong disorder.
Of course, if typical fluctuations of the
effective N\`eel temperature due to disorder are large compared to the
distance of the quantum critical point (Harris criterium), disorder
effects are important. Perhaps more importantly,
even at some distance form the QCP, in the
non-magnetic phase, a rare configuration of impurities can lead to a
small magnetically ordered region. Such exponentially rare
effects may 
dominate some of the thermodynamic properties of the system in a
finite region around the QCP \cite{neto} (Fig.~\ref{scen3});
e.g. $c_V/T \propto \chi \propto T^{-1+\lambda}$, where the exponent
$\lambda<1$ should vary continuously as a function of the distance of
the QCP.  Several strongly disordered uranium compounds can be fitted
to this scenario \cite{neto} at least over some range. An interesting
systematic approach to this problem was suggested by Narayanan {\it et
al} \cite{belitz}, 
who studied the transition in a double $\epsilon$ expansion in
$4+\epsilon$ space and $\epsilon_t$ time dimensions. They find that
rare events are indeed relevant in the renormalization group (RG)
sense if the number of order parameter components is smaller than $4$.
The flow to strong disorder in the RG equations can be interpreted as
the signature of the onset of a Griffiths phase, but also  other
strong-coupling fixed points might be possible. It would be interesting
to investigate whether the Kondo effect, which is not captured in an
expansion around $0$ time dimension will qualitatively change the RG flow.

\section{Resistivity}

In the following we will concentrate on the transport properties
within the first scenario. We consider a situation where the influence
of disorder on the magnetism is small, but we will keep track of the
effects of weak impurity scattering.

We propose \cite{roschRes} to study the interplay of isotropic elastic
scattering and anisotropic scattering from spin waves within a
semi-classical Boltzmann equation treatment of electrons interacting
with spin-fluctuations and impurities.  We use the fact that the
former are described by a theory above the upper critical dimension
\cite{hertz} and, due to the ohmic damping of the magnetic
excitations in a metal, is characterized by a dynamical exponent
$z=2$.  As a result, the spin-fluctuation spectrum in the paramagnetic
phase can be modeled by \cite{hertz,hlubina,moriya,2dim}
\begin{eqnarray} \label{chi}
\chi_{\vec{q}}(\w)\approx 
\frac{1}{1/(q_0 \xi)^2+(\vec{q}\pm \vec{Q})^2/q_0^2- i \w/\Gamma},
\end{eqnarray}
where $q_0$ and $\Gamma$ are characteristic momentum and energy
scales; $\Q$ is the ordering wave vector in the AFM phase and $\xi$ is
the AFM correlation length which is near the QCP of the form
\cite{hertz}
\begin{equation} 1/(q_0 \xi)^2=r+c
(T/\Gamma)^{3/2} \end{equation} $r$ measures the distance from the
quantum critical point and is e.g.  proportional to $p-p_c$ in an
experiment tuned by pressure $p$.  For the purposes of our numerical
calculations we set $c=1$, even at the QCP for $r=0$ the temperature
dependence of $\xi$ does not influence the low temperature
resistivity.

Our starting point is the Boltzmann equation with a quasi-particle
distribution function
$f_{\vec{k}}=f^0_{\vec{k}}-\Phi_{\vec{k}} (\partial
f^0_{\vec{k}}/\partial \epsilon_{\vec{k}})$ linearized around
the Fermi distribution $f^0_{\vec{k}}$ with a  collision term
\begin{eqnarray} 
\left.\frac{ \partial f_{\vec{k}}}{\partial t}\right|_{\text{coll}} \!\!\!
&=&  
\sum_{\vec{k}'} \frac{f^0_{\vec{k}'} (1-f^0_{\vec{k}})}{T} 
(\Phi_{\vec{k}}-\Phi_{\vec{k}'}) \label{coll}\\
&& \hspace{-1.5cm} \times
\left[ 
g_{\text{imp}}^2 
\delta(\epsilon_{\vec{k}}-\epsilon_{\vec{k}'})+\frac{2 g_S^2}{\Gamma}
n^0_{\epsilon_{\vec{k}}\!-\!\epsilon_{\vec{k}`}}
\IM \chi_{\vec{k}-\vec{k}'}(\epsilon_{\vec{k}}\!-\!\epsilon_{\vec{k}`})
\right] \nonumber
\end{eqnarray}
Here $g_{\text{imp}}^2$ and $g_S^2$ are transition rates for isotropic
impurity scattering and inelastic scattering from spin fluctuations,
respectively, and $n^0_{\w}$ is the Bose function.  In the following
we measure the resistivity in units of $\rho_M=3 \hbar g_S^2/(\pi e^2
v_F^2)$ and express our results in terms of the dimensionless elastic
scattering rate $x=\pi g_{\text{imp}}^2/(2 g_S^2)$ \cite{misprint},
the reduced temperature $t=T/\Gamma$ and the effective distance from
the critical point $r$ (Eq.~\ref{chi}).  In these units the residual
resistivity $\rho_0$ is given by $\rho_0=x \rho_M$.  Eq. (\ref{coll})
tacitly assumes that the spin-fluctuations stay in equilibrium, an
approximation valid if the spin fluctuations can loose their momentum
effectively by Umklapp or impurity scattering.

At $T=0$ the scattering is purely elastic and the
quasiparticle distribution function is given by $\Phi^0_\kk \equiv
\Phi_\kk(T=0)= e \vec{E} \vec{v}_\kk/(g_{\text{imp}}^2 N_F)$.  
For $T>0$, the AFM spin fluctuations are strongest near
the $\Q$ vectors in reciprocal space where the AFM order develops
\cite{2dim}. 
 Accordingly, the electrons with an energy
$\epsilon_{\kk} \approx \mu$ will scatter strongly from the spin
fluctuations near ``hot lines'' on the Fermi surface where
$\epsilon_{\kk} \approx \epsilon_{\kk+\Q}\approx \mu$ (right part of
Fig.~\ref{scen1}).
For sufficiently small temperatures,
 the distribution function will change only in
the neighborhood of the hot lines
 as long as the scattering rates in
the ``cold regions'' (Fig.~\ref{scen1}) are dominated by
impurities, i.e. for $t^2 \ll x$. The strong scattering due to
spin-fluctuations will equilibrate the points $\kk$ and ${\kk \pm \Q}$
along the hot lines on the Fermi surface   and the non-equilibrium 
distribution function is written in the form
$\Phi_\kk \approx (1-p_\kk) \Phi_{\kk}^0+ p_\kk \Phi_{\kk\pm \Q}^0$, where
$p_\kk$ approaches $1/2$ near the hot lines and vanishes further away. It
is helpful to use a coordinate system, where vectors on the Fermi
surface $\kk=\kk_H+\kk_{\perp}$ are split up in a vector $\kk_H$
on the hot
line and a perpendicular vector $\kk_{\perp} \perp \kk_H,\vec{v}_\kk$. 
 After rescaling the momenta using
$\kk=\kappaV q_0 \sqrt{t} $ the Boltzmann equation in the limit
$t<\sqrt{x}$ can  be rewritten in the following way
\begin{eqnarray}
t \int& d\kappaV` & (p_\kappaV+p_{\kk_H\pm \Q+\kappaV'}-1) 
I(r/t+M_{\kappaV \kappaV'}) \nonumber\\
&& +c_{\kk_H} x p_\kappaV = 0  \label{bolzScaling}
\end{eqnarray}
\begin{equation}
I(y)=\int_0^{\infty} dz z^2 n^0(z)(1+n^0(z))/(y^2+z^2),
\end{equation}
where $\kappaV'=\kappaV'_{\|}+\kappaV'_{\perp}$,
$M_{\kappaV \kappaV'}={\kappa'_{\|}}^2+
(\kappaV'_{\perp}-\kappaV_{\perp})^2$
and $c_{\kk_H}= (2 \pi)^3 N_F v_{\kk_H+\Q}/(\pi q_0^2)$. 
From (\ref{bolzScaling}) it is clear that only the combinations 
$t/x$ and $t/r$ determine $p_{\kappaV}$. Accordingly, the resistivity
obeys a scaling relation for $t \ll \sqrt{x} \ll 1$, $r\ll1$:
\begin{figure}[t]
  \centering
 \includegraphics[width=0.75 \linewidth]{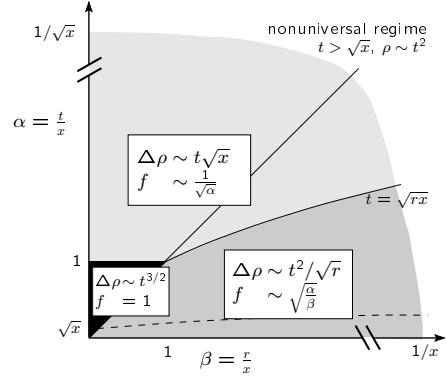}
\caption[]{In the scaling limit $t,x,r \to 0$, $t/x,r/x \to const.$
the resistivity is ``universal'' $\Delta \rho/\rho_M =
 t^{3/2} f(t/x,r/x)$ for $t<\sqrt{x}, r<1$, 
where $t\propto T$, $x\propto \rho_0$ and $r\propto p-p_c$
measure the temperature, the amount of disorder and the distance from the
QCP. The plot shows the qualitative behavior of the scaling function $f$ in
the various regimes. For $\min[x, \sqrt{r x}<t<\sqrt{x}]$
the resistivity rises  linearly with temperature. Thermodynamic
quantities show a crossover to Fermi liquid behavior at the scale $t=r$,
while in transport a $T^2$ behavior is recovered at a much smaller
scale $\min[x, \sqrt{r x}]$. The dashed line serves as a reminder
that at lowest temperatures 
effects which are not included in our approach become important, e.g.
interference effects of disorder and interactions \cite{altshuler}
or a disorder induced change of the spin fluctuation spectrum
\cite{belitz}.
}
\label{scaleRho}
\end{figure}
\begin{eqnarray}
\frac{\Delta \rho}{\rho_M}&\approx& t^{\frac{d}{2}} f\!\left(\frac{t^{
\frac{d-1}{2}}}{x},
\frac{r^{\frac{d-1}{2}}}{x}\right) \\
&\sim& \left\{ 
\begin{array}{ll}
t^{d/2}& , r<t<x^{2/d-1} \\
t^{\frac{2}{5-d}} x^{\frac{4-d}{5-d}} &, 
\text{max}[x^{\frac{2}{d-1}},\sqrt{x} r^{\frac{5-d}{4}}]<t<\sqrt{x} \\
t^2/r^{2-\frac{d}{2}} &, t<\text{min}[r,\sqrt{x} r^{\frac{5-d}{4}}]
\end{array}
\right. \nonumber
\end{eqnarray}
where for completeness we have given the result in $d$ dimension. Note,
however, that the Boltzmann equation approach presumably breaks  down in
$d=2$ as the ``hot'' electrons acquire a lifetime $\propto \sqrt{T}$.
An overview of the behavior of $\Delta \rho$ in $d=3$ 
is given in Fig.~\ref{scaleRho}.

The asymptotic behavior can be easily extracted from (\ref{bolzScaling})
by realizing, that $p_\kk$ varies smoothly parallel to the hot lines,
allowing us to replace $p_{\kk_H \pm \Q+ \kappaV'}
\equiv p_{\kappa'_\| \kappa'_\perp }$ by 
$p_{\kappa_\| \kappa'_\perp }$ in  (\ref{bolzScaling}) and to perform
the $\kappa'_\|$  integration.
For dominant impurity scattering, $t < \min[x,\sqrt{r x}]$, 
one can use a perturbative expansion
in the first term in (\ref{bolzScaling}) and obtains
\begin{eqnarray}
\frac{\Delta \rho}{\rho_M}&\approx&
t^{3/2} h\!\left(\frac{t}{r}\right)  \frac{3 q_0^3}{8 \pi^4 (v_F N_F)^2 }
\nonumber \\ \label{rhoDis}
&& \times \sum_i 
\oint_{i} d \kk_{\|} \frac{((\vec{v}_{\kk_{\|}}-\vec{v}_{\kk_{\|} \pm \Q_i}) 
\vec{n})^2}
{|(\vec{v}_{\kk_{\|}} \times \vec{v}_{\kk_{\|}\pm \Q_i}) \hat{\vec{e}}_\||} 
\label{rhoImp} \\
h\left(\frac{t}{r}\right) &=& \frac{\pi}{2} \int_0^{\infty}
 y n^0(y)(1+n^0(y))  \IM \sqrt{\frac{r}{t}+ i y} \nonumber \\
&\approx& \left\{ \begin{array}{ll}
\frac{\pi \zeta(3/2) \Gamma(5/2)}{2 \sqrt{2}} 
 &, t >r  \\[.3cm]
\frac{\pi^3}{12}\sqrt{t/r}
&, t< r \Gamma 
\end{array} \right.
\end{eqnarray}
where $\hat{\vec{e}}_\|$ is a unit vector parallel to the hot line and
$\sum_i$ sums over all hot lines. Only in this dirty limit, the
commonly used \cite{moriya} approximation is valid to calculate the
resistivity from the average of the scattering rate.

\begin{figure}[t]
  \centering
 \includegraphics[width=0.7 \linewidth]{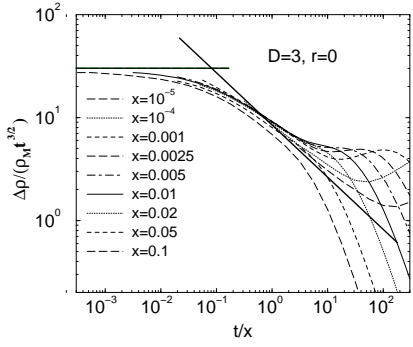}
\caption[]{Scaling plot of $f(t/x,0)=\Delta \rho(t/x)/t^{3/2}$ 
at the QCP ($r=0$) using a numerical solution of the Boltzmann equations
for various amounts of disorder $x$.
The straight solid lines show the limiting cases $\Delta \rho \propto t^{3/2}$
for $t \ll x$ and $\Delta \rho \propto t$ for $x \ll t \ll \sqrt{x}$.
Deviations from scaling are large for $t/x>1/\sqrt{x}$.}
\label{figScaling}
\end{figure}

In the limit $\sqrt{x}>t>\min[x,\sqrt{r x}]$,  the resistivity rises
linearly with temperature
\begin{eqnarray}
\frac{\Delta \rho}{\rho_M}\approx&& t\sqrt{x} 
\frac{q_0^2  \sqrt{\pi/3}}{2 v_F^2 (2 \pi)^3 N_F^{3/2}}  \sum_{i} \oint_{i} 
d \kk_{\|} \biggl[ \nonumber \\
&&  \frac{((\vec{v}_{\kk_{\|}}-\vec{v}_{\kk_{\|} \pm \Q_i}) 
\vec{n})^2}
{|(\vec{v}_{\kk_{\|}} \times \vec{v}_{\kk_{\|}\pm \Q_i}) \hat{\vec{e}}_\||} 
\frac{\vec{v}_{\kk_{\|} \pm \Q_i}^{5/4}}{\vec{v}_{\kk_{\|}}^{3/4}} 
S_{\vec{v}_{\kk_{\|} \pm \Q_i}/\vec{v}_{\kk_{\|}}}^
{|\hat{\vec{v}}_{\kk_{\|}\pm \Q_i} \hat{\vec{v}}_{\kk_{\|}} |}\biggr] \nonumber\end{eqnarray}
where the only slightly varying function $S_{\alpha}^a=\int_{-\infty}^{\infty}
 d \kappa p_\kappa(a,\alpha) \approx \pi$ is calculated from
the solution of the two coupled one-dimensional integral equations
\begin{eqnarray}
p_\kappa+\int d \kappa'  \frac{(p_\kappa+
\tilde{p}_{\kappa'}-1)(1-a^2)}{(\kappa^2/\alpha+{\kappa'}^2 \alpha-
2 \kappa \kappa' a)^{3/2}}&=&0\nonumber \\
\tilde{p}_\kappa+\int d \kappa'  \frac{(\tilde{p}_\kappa+
p_{\kappa'}-1)(1-a^2)}{(\kappa^2 \alpha+{\kappa'}^2/\alpha-
2 \kappa \kappa' a)^{3/2}}&=&0
\end{eqnarray}
which in the limit $a\to \pm 1$ are solved by 
$p_\kappa=1/((1+\alpha)+\kappa^2/2 \sqrt{\alpha})$.

For $1 \gg t \gg \sqrt{x}$ the resistivity is much less
universal and
as pointed out by Hlubina and Rice \cite{hlubina}
 $\rho/\rho_M \approx t^2$, where the prefactor depends on all details
of the scattering processes in the ``cold regions''. The full scaling
curve for $r=0$  and the deviations
from scaling  are shown
Fig.~\ref{figScaling}.

\section{Conclusions}

A main qualitative difference of the SDW scenario (Fig.~\ref{scen1})
discussed above and a scenario, where the Kondo effect breaks down at
the transition (Fig.~\ref{scen2}), is that in the first case only a
small fraction of the quasi particles near the ``hot lines'' is
strongly scattered by the critical fluctuations (Fig.~\ref{scen1})
while in the second case the full Fermi surface is affected.

At first glance, there is a simple way to distinguish these two
situations \cite{coleman}: for an ultra-clean Fermi liquid interacting
with spin-fluctuations the quasi particles with the longest lifetime
short-circuit those with a strong scattering rate and determine the
resistivity.  Accordingly, one expects a Fermi liquid like $T^2$
resistivity in a system without disorder within the SDW scenario
\cite{hlubina} (the limit $ x \ll t^2 \ll 1$ discussed above).  This would
suggest, that a low-temperature resistivity of the form $T^\alpha$
with $\alpha<2$ in a clean system signalizes a total breakdown of the
Fermi liquid.  Our calculations, however, show that the ultra-clean
limit is irrelevant even for samples with a RRR of $10^3$ or $10^4$
\cite{roschRes} because very weak impurity scattering leads to the
qualitatively different resistivity discussed above.

\begin{figure}[t]
  \centering
 \includegraphics[width=\linewidth]{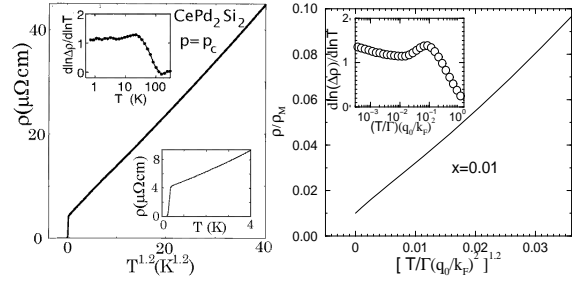}
\caption[]{Resistivity as a function of $T^{1.2}$ 
  in CePd$_2$Si$_2$ taken from \cite{julian} (left figure)
 compared
  to our calculation (right figure) for $x=0.01$. 
  The insets show the corresponding
  logarithmic derivative of $\rho(T)-\rho(0)$.
 Below $\approx 400$mK CePd$_2$Si$_2$ is
  superconducting (lower inset).
  Note the offset of the line $T=0$ in
  both plots.}
\label{figExperiment}
\end{figure}

Our results imply that in all situations
where two scattering mechanisms compete, one
anisotropic and affecting only parts of the Fermi surface, the other
isotropic, the transport is quite subtle and  the interplay of the
two mechanisms can lead to qualitatively new effects.
For example, Mathiessen's rule is not valid in this regime.

Taking this into account, it seems that quite a number of heavy
Fermion systems are good candidates to be described within our first
scenario, a SDW transition in a Fermi liquid (e.g. CePd$_2$Si$_2$,
CeNi$_2$Ge$_2$, CeIn$_3$, CeCu$_2$Si$_2$, CeNiGa$_2$
\cite{julian,gegenwart}).  The qualitative trends are well described
by the theory: $\Delta \rho \propto T^{1.5}$ in the more dirty systems
and $\Delta \rho \propto T^\alpha$ with $\alpha$ close to $1$ in the
cleaner systems. Also the cleaner systems seem to show anomalous
behavior in the resistivity (but not in thermodynamics) even at some
distance away from the QCP, consistent with our results (see
Fig.~\ref{scaleRho}). In Fig.~\ref{figExperiment} the resistivity of a
quite clean sample of CePd$_2$Si$_2$ \cite{julian} is compared to a
typical solution of the Boltzmann equation for $x=0.01$ (see
\cite{roschRes} for the precise model used).  It is important to point
out, that not only a similar effective exponent shows up in both
theory and experiment, but that it is also observed over a similar
range $0.1 \rho_0 < \Delta \rho(T) <10 \rho_0$. Certainly, it has to
be checked carefully whether and to what extend other predictions of
the SDW scenario are observed experimentally.  E.g. the predicted
\cite{hertz} pressure dependence of the N\'eel temperature $T_N
\propto (p_c-p)^{2/3}$ appears to disagree with the experimentally
observed linear behavior in CePd$_2$Si$_2$ \cite{julian}.

The best test of our prediction would be a careful comparision of
samples of different quality e.g. using a scaling analysis as shown in
Fig.~\ref{figScaling}. Certainly, more work is necessary to
identify the origin of the non-Fermi-liquid behavior in systems like
CeCu$_{6-x}$Au$_x$ and to clarify the role of strong disorder.

% Acknowledgements are optional.
\begin{ack}
\vspace{-.2cm}
I would like to thank   A.V. Chubukov,
P.~Coleman, M.~Grosche, H.v.~L\"ohneysen,
 C.~Pfleiderer, A.E.~Ruckenstein and
 P.~W\"olfle for discussions and the
 A.v.~Humboldt
foundation and the Center for Materials Theory at Rutgers University
for support.

\end{ack}
\vspace{-.2cm}

% References


\begin{thebibliography}{9}
\vspace{-.2cm}

\bibitem{loehneysen}  H.v. L\"ohneysen {\it et al.},
Phys. Rev. Lett. {\bf 72}, 159 (1994); H. v. L\"ohneysen, J. Phys.: Cond. Matt. {\bf 8}, 9689
%(1996) and refs. therein.

\bibitem{julian} S. R. Julian, {\it et al.}, J. Phys. Cond. Mat. {\bf 8},
9675 (1996); F.M.~Grosche {\it et al.}, preprint, cond-mat/9812133;
 N. D. Mathur, {\it et al.}, Nature {\bf {394}}, 39 (1998).
%F. M. GROSCHE, S. R. JULIAN, I. R. WALKER, D. M. FREYE,
%R. K. W. HASELWIMMER & G. G. LONZARICH

\bibitem{gegenwart} P. Gegenwart {\it et al.},  Phys. Rev. Lett. {\bf 81}, 1501 (1998); R. Hauser {\it el al}, JMMM {\bf 177-181}, 292 (1998);
 R. Heuser, E.-W. Scheidt, T. Schreiner, and G.R. Stewart
, Phys. Rev. B {\bf 58}, R15959 (1998).


\bibitem{super}
In very clean samples of CePd$_2$Si$_2$, CeIn$_3$
and CeGe$_2$Si$_2$ also a superconducting phase has been reported 
close to the AFM QCP \cite{julian}.

\bibitem{doniach} S. Doniach, Physica B {\bf 91}, 231 (1977).

 \bibitem{hertz} J. A. Hertz,
Phys. Rev. B {\bf 14}, 1165 (1976);
 A. J. Millis, Phys. Rev. B {\bf 48}, 7183 (1993).


\bibitem{hlubina} R. Hlubina and T. M. Rice, Phys. Rev. B {\bf 51},
 9253 (1995).
 
\bibitem{roschRes} A. Rosch, 
Phys. Rev. Lett {\bf 82}, 4280 (1999).

\bibitem{si} Q. Si, J. L. Smith, and K. Ingersent,
cond-mat/9905006. Q. Si and J. L. Smith, cond-mat 9903083.

\bibitem{schroeder}
A. Schr\"oder {\it et al.},
%G. Aeppli, E. Bucher, R. Ramazashvili and P. Coleman, 
Phys. Rev. Lett. {\bf 80}, 5623 (1998);
%\bibitem{stockert97} O. Stockert, H. v. L\"ohneysen, A. Schr\"oder, M.
%Loewenhaupt, N. Pyka, P. L. Gammel and U. Yaron, Physica B {\bf 230}, 247
%(1997).

\bibitem{coleman} 
P.~Coleman, preprint, cond-mat/9809436.

\bibitem{altshuler} B.L.~Altshuler and A.G. Arononv, in {\it Electron-Electron
Interaction in Disordered Systems}, edited by A.L. Efros and M. Pollak
(North Holland, Amsterdam, 1985).

\bibitem{belitz} R. Narayanan, T. Vojta, D. Belitz, T. R. Kirkpatrick,
cond-mat/9905047. T. R. Kirkpatrick and D. Belitz, Phys. Rev. Lett. {\bf 76}, 2571, (1996); {\it ibid.} {\bf 78}, 1197 (1997).

\bibitem{neto} A. H. Castro Neto, G. Castilla, and  B.A. Jones, Phys. Rev. Lett. {\bf 81} 3531 (1998) ; M. C. de Andrade {\it et al}, Phys. Rev. Lett.
 {\bf 81} 5620 (1998);

\bibitem{moriya} T. Moriya and T. Takimoto, J. Phys. Soc. Jpn. {\bf
    64}, 960 (1995).
% S. Kambe {\it et al.},
% S. Raymond, J. McDonough, L.P. Regnault, J. Floquet, P. Lejay, and P. Haen,
%J. Phys. Soc. Japam {\bf 65} (1996) 3294.

\bibitem{2dim} We assume a three dimensional nature of the spin
fluctuations and do not consider strongly anisotropic cases as have
been observed in CeCu$_{5.9}$Au$_{0.1}$ \cite{stockert}.

\bibitem{stockert}
A. Rosch {\it et al.},
% A. Schr\"oder, O. Stockert, H.v. L\"ohneysen,
Phys. Rev. Lett. {\bf 79}, 159 (1997); O.~Stockert {\it et al.}, 
Phys. Rev. Lett. {\bf 80}, 5627 (1998);
% O. Stockert, A. Schr\"oder, H. v. L\"ohneysen, A. Rosch,
%N. Pyka, M. Loewenhaupt

%\bibitem{danger} The interaction $U$ is ''dan\-ger\- 1 
%ously'' irrelevant, it
%determines the temperature-dependence of the correlation length,
%$1/\xi^2 \propto U T^{3/2}$. 

\bibitem{misprint} This definition differs from the one given in
 \cite{roschRes} by a factor $\pi$ correcting a misprint.


\end{thebibliography}
\end{document}